\documentclass[twocolumn,pra,aps,showpacs]{revtex4}

\usepackage{amsmath}
\usepackage{amssymb}
\usepackage[dvips]{graphicx}
\usepackage{pstricks}
\usepackage{pst-node}
\usepackage{pst-plot}

\newcommand{\qed}{\hspace*{\fill}$\square$}

\newcommand{\be}{\begin{equation}}
\newcommand{\ee}{\end{equation}}

 \newcommand{\sset}[1]{ \{#1\} }
 \newcommand{\half}{\frac 1 2}

 \newcommand{\ket}[1]{|#1\rangle}

 \newcommand{\qubit}{\mathcal{H}_2}
 \newcommand{\qubits}[1]{\mathcal{H}_2^{\otimes #1}}
 \newcommand{\Pauli}[1]{\mathbf P_{#1}}

\begin{document}

\title[Short Title]{Optimal Resources for
Topological 2D Stabilizer Codes: Comparative Study}

\author{H. Bombin and M.A. Martin-Delgado}
\affiliation{ Departamento de F\'{\i}sica Te\'orica I, Universidad
Complutense, 28040. Madrid, Spain. }

\begin{abstract}
We study the resources needed to construct topological 2D stabilizer
codes as a way to estimate in part their efficiency and this leads
us to perform a comparative study of surface codes and color codes.
This study clarifies the similarities and differences between these
two types of stabilizer codes. We compute the error correcting rate
$C:=n/d^2$ for surface codes $C_s$ and color codes $C_c$ in several
instances. On the torus, typical values are $C_s=2$ and $C_c=3/2$,
but we find that the optimal values are $C_s=1$ and $C_c=9/8$. For
planar codes, a typical value is $C_s=2$, while we find that the
optimal values are $C_s=1$ and $C_c=3/4$. In general, a color code
encodes twice as much logical qubits as a surface code does.
\end{abstract}


\maketitle

\section{Introduction}
\label{sect_intro}

Decoherence of quantum states is one of the main reasons why we have
not achieved so far many of the impressive results predicted by
quantum information theory. Battling decoherence has become a very
important issue in this field. Devising new strategies to deal with
decoherence effects is equally important. One of these strategies,
``the topological way'', relies on quantum states endowed with a
robustness arising when they are embedded into certain Hilbert
spaces that exhibit topological protection \cite{kitaev97}.

Quantum error correction has provided us with definite techniques to
do error correction on quantum states belonging to quantum codes \cite{shor95}, \cite{steane96a}.
A suitable formalism to study quantum error correction codes is the
stabilizer formalism \cite{gottesman96}. In fact, a topological quantum code is a special type
of stabilizer code \cite{kitaev97}, as will be discussed in sect.\ref{sect_topo2D}.
It is a reservoir of states that are intrinsically robust against decoherence due to
the encoding of information in the topology of the system.

From the point of view of quantum computation, a quantum error
correcting code is a quantum memory \cite{dennis_etal02},
\cite{optimalGraphs}, \cite{homologicalerror}. Thus, a topological
code amounts to a quantum memory with topological protection and it
can be endowed with extra computational capabilities under certain
circumstances \cite{tetraUQC}, \cite{rhg06}, \cite{rhg07}. This
property is rather convenient since one of the advantages of
topological codes is that they give rise to self-protecting quantum
memories. This means that the system has physical resources to
perform the error correction process by itself, once a local error
pops up in the code. In other words, the process of error correction
is done by hardware means, not by software operations such as
checking and measuring an error syndrome.

The physical mechanism that underlies a topological quantum code is called a topological order
\cite{wenniu90}, \cite{wen90}, \cite{wenbook04}. This is a new type of quantum phase for matter.
In a topological order there exists ground state degeneracy
without breaking any symmetry, in sharp contrast with more standard phases based on the
spontaneous symmetry breaking mechanism. This degeneracy has a topological origin.
Thus, topological orders deviate significantly from more standard orders treated
within the Landau symmetry-breaking theory \cite{levinwen05}, \cite{topo3D}, \cite{fzx07}, \cite{hl06}.

Topological protection is very appealing and has many virtues, but
there are also difficulties to implement it in practice. This is
currently an active and broad area. We shall not be concerned with
experimental realizations of topological codes here.

Our main interest is to analyze the resources needed for their
construction and the optimality of those resources.
In doing so, we shall perform a very illustrative comparative
study of the similarities and differences between the
main examples of topological stabilizer codes, namely,
surface codes \cite{kitaev97} and color codes \cite{topologicalclifford}.

This paper is organized as follows:
 in Sect.\ref{sect_topo2D} we introduce the surface codes in a slightly different manner
 than the usual one \cite{kitaev97}, but otherwise equivalent. We do so because its comparison
 with color codes \cite{topologicalclifford} is more transparent in this way. Color codes
are constructed with certain two-dimensional complexes, 2-colexes,
introduced in \cite{topo3D}. We point out the shortcoming of the
surfaces codes with respect to color codes as far as the
implementation of a variety of important transversal quantum logic
gates belonging to the Clifford group. Another advantage of color
codes is that they encode twice the number of logical qubits than
surface codes do. In Sect.\ref{sect_efficiency} we introduce the
notion of error correcting rate $C$ for any code. It gives
information about how good a code is for error correction when the
number of physical qubits $n$ is increased. We compute this scaling
both for surface codes and color codes with the topology of a torus
and a plane, which are the most important examples of topologies in
2D for practical reasons. Moreover, we compute the optimal values of
this figure of merit $C$ for those topological 2D stabilizer codes.
Sect.\ref{sect_conclusions} is devoted to conclusions.

\section{Topological 2D Stabilizer Codes}
\label{sect_topo2D}

We start introducing the notion of a stabilizer quantum error
correcting code \cite{gottesman96}. Let $X$, $Y$ and $Z$ denote the
usual Pauli matrices, which act on the space $\qubit$ of a single
qubit. A Pauli operator $p_n$ of length $n$ is any tensor product of
the form \be p_n:=\bigotimes_{i=1}^n \sigma_i, \ \  \sigma_i \in
\sset{I,X,Y,Z}. \ee
\begin{figure}
 \includegraphics[width=7cm]{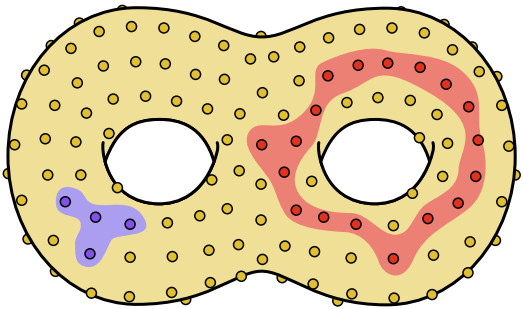}
 \caption{
 A 2-torus is an example of a topological space in which a
 topological stabilizer code can be constructed. Here the dots
 represent the qubits pinned down onto the surface. The small blue
 area is the support of a local generator of the stabilizer $\mathcal S$
 \eqref{stabilizer_condition}. The
 big red area, which cannot be deformed to a point, is the support
 of an undetectable error (in $\mathcal Z - \mathcal S$).
 }
 \label{figura_topocodes}
\end{figure}

The closure of such operators as a group is the Pauli group
$\Pauli n$. Given an Abelian subgroup $\mathcal S\subset\Pauli n$,
an stabilizer code of length $n$ is the subspace $\mathcal C\subset\qubits n$
formed by those vectors $\ket{\phi}$ with eigenvalue 1 for
any element  $s\in \mathcal S$,
\be
s \ket{\phi} = \ket{\phi}.
\label{stabilizer_condition}
\ee
Let $\mathcal Z$ be the centralizer
of $S$ in $\Pauli n$, i.e., the set of operators in $\Pauli n$
that commute with the elements of $\mathcal S$. A Pauli operator
$z\in \mathcal Z$ not contained in $S$ up to a phase leaves
$\mathcal C$ invariant and acts nontrivially in $\mathcal C$. Such
operators, when regarded as errors, are clearly undetectable. Let
the weight of an operator be the number of qubits in which it acts
nontrivially. Then the minimal length among the operators in
$\mathcal Z- \mathcal S$ is called the distance of the code.
Indeed, the code is capable of correcting a set of Pauli errors
$\mathcal E$ as long as for any $M,N\in\mathcal E$ the operator
$M^\dagger N$ is not an undetectable error. Therefore, a code of
distance $d=2t+1$ can correct all the errors of length less or
equal to $t$. Given $z \in \mathcal Z$ and $s \in\mathcal S$, $z$
and $zs$ act equally in $\mathcal C$. Then choosing suitably among
the equivalence classes of $\mathcal Z/\mathcal S$, we can find a
Pauli operator basis for the encoded qubits.

Topological stabilizer codes can be roughly defined as stabilizer
codes in which the generators of $\mathcal S$ can be chosen to be
local and undetectable errors have a support that is topologically
nontrivial, as shown in Fig.~\ref{figura_topocodes}. We are assuming
that the physical qubits that make up the stabilizer code are placed
in certain topological space. In particular, we will only consider
codes placed onto two-dimensional surfaces. One of the ideas behind
topological stabilizer codes is that the locality of the generators
is something very advantageous in order to perform error correction.
Another important idea is that of self-protected quantum memories,
something that we will touch upon later.

\begin{figure}
 \includegraphics[width=8cm]{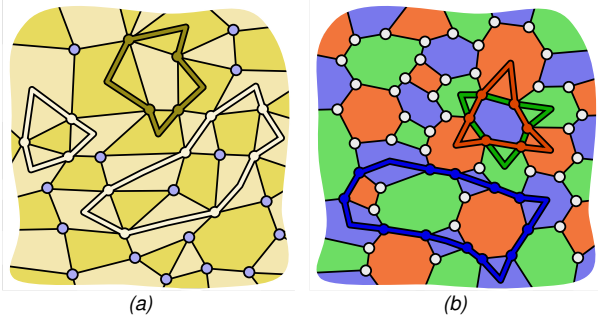}
 \caption{
 Typical lattices for both kinds of 2-D topological stabilizer
 codes. In both cases qubits are represented as circles placed at
 sites. (a) A piece of a surface code \cite{kitaev97}. Dark plaquettes have $B_p^X$
 stabilizers attached, while light plaquettes have $B_p^Z$ stabilizers attached. In both
 cases these operators correspond to closed strings. For example, an $X$-string
 operator which is the product of three dark plaquette operators is
 shown. (b) A piece of a color code \cite{topologicalclifford}. All plaquettes have $B_p^X$
 and $B_p^Z$ stabilizers attached, which can be both visualized as two
 strings of different colors. In the case of a blue plaquette, its operators
 can be considered either as red or as green strings. For example, a blue string
 operator which is the product of two red and one green plaquette
 operators is shown.
 }
 \label{figura_codes}
\end{figure}

The first example of topological stabilizer codes were toric
codes \cite{kitaev97}, in which the qubits are placed in a torus. More generally,
other surfaces and arbitrary lattices on them can be considered,
and the resulting codes were termed in general surface codes \cite{bravyikitaev98},
\cite{dennis_etal02}. We will introduce here surface codes in a way that differs slightly
from the original one but is absolutely equivalent. Consider any
tetravalent lattice \cite{tetravalent} with bi-colorable plaquettes, such as the one
shown in Fig.~\ref{figura_codes}(a). The plaquettes are split
into two sets, which we label as dark and light sets of plaquettes.
A surface code can be obtained from such a lattice
placing a qubit at each of its sites and choosing suitable
plaquette operators. In general, given a plaquette $p$ we define
the plaquette operators
\be
B_p^\sigma:= \bigotimes_i \sigma^{s_p(i)}, \qquad \sigma = X,Z,
\ee
where the product is over all the sites and $s_p(i)$ equals one
for sites belonging to $p$ and is zero otherwise. In the case of
surface codes, the generators of $\mathcal S$ are $B_p^X$ for dark
plaquettes and $B_p^Z$ for light plaquettes. Note how all these
operators commute, thus generating an Abelian subgroup.
The encoded states $\ket\phi$ satisfy the
conditions
\begin{eqnarray}
\forall p \in P_D\qquad B_p^X\ket\phi &=& \ket \phi, \\
\forall p \in P_L\qquad B_p^Z\ket\phi &=& \ket \phi,
\end{eqnarray}
where $P_D$ and $P_L$ are respectively the sets of dark and light
plaquettes.

Let a $Z$-operator ($X$-operator) be any tensor product of $Z$s
($X$s) and $I$s. Then any $Z$-operator ($X$-operator) can be
visualized as a string that connects dark (light) plaquettes and
acts nontrivially on those qubits it goes through, see
Fig.~\ref{figura_codes}(a). Any light (dark) plaquette operator is a
$Z$-string ($X$-string) operator. Then the product of several
plaquette operators of the same kind is a string operator lying on
the boundary of certain area containing precisely the plaquettes,
see Fig.~\ref{figura_codes}(a). Any Pauli operator is, up to a
phase, the product of an $X$-string and a $Z$-string. In this sense,
the strings belonging to $\mathcal S$ are all boundaries. More
generally, any operator in $\mathcal Z$ is a product of closed
string operators. Closed strings are strings without endpoints, and
their importance is now clear since those closed strings which are
not boundaries make up precisely the set of undetectable errors.
From these undetectable errors we can choose a Pauli operator basis
for the encoded qubits, that is, we can choose the encoded $Z$s and
$X$s operators, acting on the logical qubits. It is customary to
denote these encoded operators as $\bar{Z}$ and $\bar{X}$ to
distinguish them from the standard operators acting on the physical
qubits instead. To this end, we note two properties:

\noindent i/ String operators of the same kind that differ only by a
boundary, that is, which are equal up to a deformation, have the
same action on encoded states.

\noindent ii/ A light and a dark string operator commute if they
cross an even number of times and anticommute otherwise. Observe
that this crossing parity is preserved by the string deformations
just mentioned.

\begin{figure}
 \includegraphics[width=8cm]{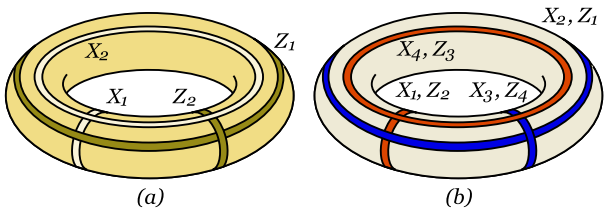}
 \caption{
 Pauli operator bases in the torus. The encoded operators correspond
 to certain strings operators. (a) For surfaces codes \cite{kitaev97}, the
 number of encoded qubits is two. (b) For color codes \cite{topologicalclifford}, the number of
 encoded qubits is four. Note that each string has two
 operators attached, one of each type.
 }
 \label{figura_bases}
\end{figure}

Taking this into account, one can obtain the desired Pauli operator
basis and find out that the number of encoded qubits is $2g$ for a
$g$-torus, that is, a sphere with $g$ `handles', see
Fig.~\ref{figura_bases}(a).

The topological nature of surface codes makes them very attractive.
In particular, the measurements required for quantum error
correction can be locally performed and involve the few qubits lying
on each plaquette. On the other hand, they are not so nice if one
intends to perform transversal operations with codes. In fact, only
the CNot gate can be performed transversally in surface codes. It
was precisely with the aim to overcome this difficulty that color
codes \cite{topologicalclifford} were devised, which are also 2D
topological stabilizer codes but allow the transversal
implementation of any operation in the Clifford group. This set of
operations is specially suited for quantum information tasks such as
quantum teleportation or entanglement distillation.

In the case of color codes \cite{topologicalclifford}, the starting point is a trivalent
lattice with tri-colorable plaquettes, see
Fig.~\ref{figura_codes}(b). We label the plaquettes as green, red
or blue. Again qubits must be placed at sites, but now for each
plaquette $p$ we have both operators $B_p^X$ and $B_p^Z$ as
generators of the stabilizer $\mathcal S$. The encoded states
$\ket\phi$ satisfy the conditions
\begin{eqnarray}
\forall p \qquad B_p^X\ket\phi &=& \ket \phi, \\
\forall p \qquad B_p^Z\ket\phi &=& \ket \phi.
\end{eqnarray}
As in the case of surface codes, string operators are essential for
the analysis of color codes. However, now the same geometrical
string can be attached to two different operators, the corresponding
$X$-string and $Z$-string. An extra labeling turns out to be
extremely useful, and so we speak of red, green and blue strings.
Blue strings connect blue plaquettes, and so on, just as $X$-strings
connect dark plaquettes in surface codes. Observe how any blue
plaquette operator, for example, can be considered both a green and
a blue string operator. Also, the product of, say, several green and
red plaquette operators is a blue string operator lying in the
boundary of certain area containing precisely the plaquettes, see
Fig.~\ref{figura_codes}(b). As in surface codes, the strings
appearing in $\mathcal S$ are all boundaries, and the strings
appearing in $\mathcal Z$ are closed strings. Also, those closed
string which are not boundaries comprise undetectable errors and
from these undetectable errors we can choose the encoded $\bar{Z}$s
and $\bar{X}$s. Again, we have two guiding properties:

\noindent i/ String operators of the same type ($X$ or $Z$) and color that are equal up to a
deformation have the same action on encoded states.

\noindent ii/ String operators commute one another, unless they cross an odd number of times
and have different color and type.

Taking this into account one can obtain a Pauli operator basis and
find out that the number of encoded qubits is $4g$ for a $g$-torus,
that is, two times the number of encoded qubits in surface codes,
see Fig.~\ref{figura_bases}(b). It is customary to denote a quantum
error correcting code made with $n$ physical qubits, encoding $k$
logical qubits and with distance $d$ as $[[n,k,d]]$. With this
notation we have that for a fixed surface topology \be k_c = 2 k_s,
\label{encoding} \ee where the subscript $c$ stands for color codes
and $s$ for surface codes. Thus, we see that color codes are more
efficient than surface codes as far as the number of encoded qubits
is concerned. However, we may wonder whether this doubling of
logical qubits has been achieved at the expense of introducing a
bigger number of physical qubits $n$ or whether it affects the
correcting capabilities $d$ of the code.

\begin{figure}
 \includegraphics[width=8cm]{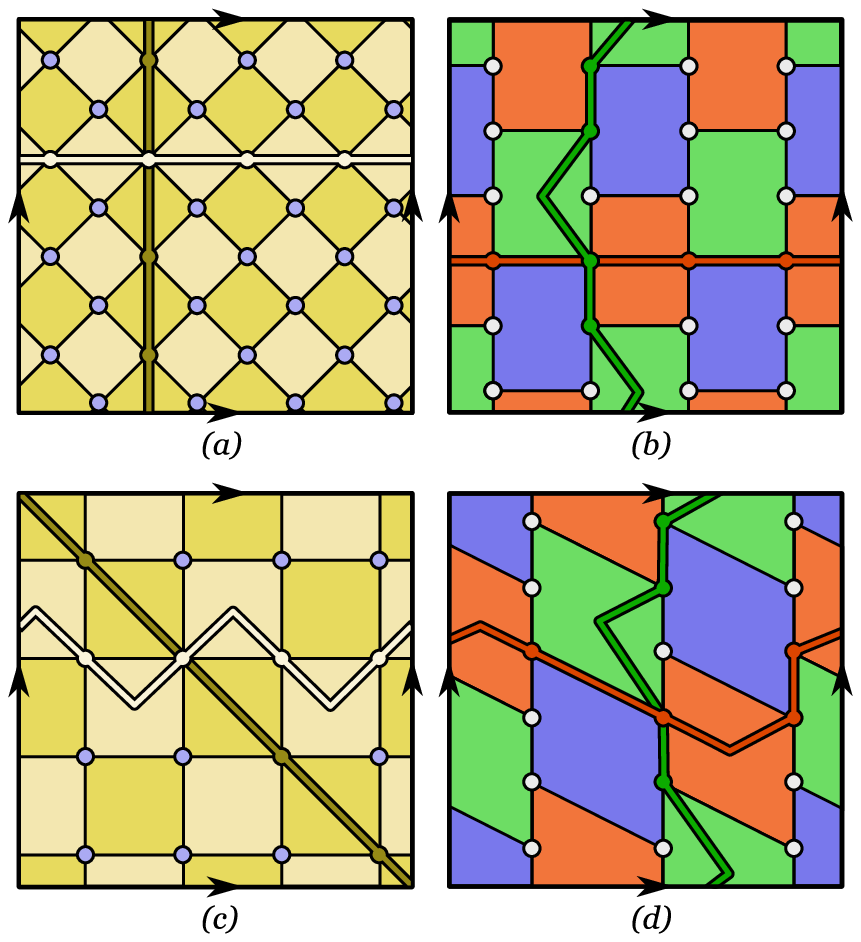}
 \caption{
 Examples of regular codes in the torus with distance $d=4$. Some
 nontrivial strings of minimal length are displayed.
  (a) In the surface code \cite{kitaev97}, plaquettes are squares and the number of
  physical qubits is $n=32$. (b) For the color code \cite{topologicalclifford}, plaquettes are
  hexagons and $n=24$. (c) An optimal regular surface code
  reduces the number of qubits to $n=16$. (d) An optimal regular color code with
  $n=18$.
 }
 \label{figura_regulares}
\end{figure}

\section{Efficiency of Topological Codes}
\label{sect_efficiency}

To answer those questions, we need to study how efficient 2-D
topological stabilizer codes are in terms of the number of  qubits
required with respect to the distance of the code. In fact, regular
lattices in which all plaquettes have the same number of qubits are
specially relevant. More specifically, it is instructive to consider
surface and color codes obtained from regular lattices on the torus.
In this case, the plaquettes must be squares for surface codes, and
hexagons for color codes, see Fig.~\ref{figura_regulares}. Let us
consider first the family of surface codes corresponding to
Fig.~\ref{figura_regulares}(a), which was in fact the first family
of topological stabilizer codes \cite{kitaev97}. The code in the
figure has distance $d=4$, the number of physical qubits is $n=32$
and the number of encoded qubits is $k=2$. More generally, this
particular example can easily be generalized to a family of codes in
which clearly \be C_s:=\frac{n_s}{d_s^2} = 2. \ee Here, we have
defined the notion of error correcting rate $C_s$ for surface codes,
a figure of merit which allows us to compare 2-D topological
stabilizer codes. It is a measure of how the error correction
capabilities of the code scales when the number of physical qubits
is incresed. The fact that the number of required qubits scales
cuadratically with distance is not surprising and is a common
feature of all 2D topological stabilizer codes. However, the
asymptotic value of $C$ for a quantum error correction code may
vary, and the value 2 is not a particularly good one, as we shall
see. Let us consider now the color code in
Fig.~\ref{figura_regulares}(b). It encodes $k=4$ qubits, is made up
of $n=24$ qubits and its distance is $d=4$. Then we have that the
error correcting rate for this color code is \be
C_c:=\frac{n_c}{d_c^2} = \frac{3}{2}. \ee Moreover, this is just an
example of an infinite family in which this ratio is preserved, so
that apparently color codes do not only encode more qubits
\eqref{encoding}, but also require less physical qubits for a given
distance.

However, as was noted in \cite{optimalGraphs}, the surface codes
just considered are not optimal in terms of the number of physical
qubits $n$. In fact, if the optimal codes are chosen, only half of
them are really needed, as Fig.~\ref{figura_regulares}(c)
illustrates. Thus, for optimal regular surface codes in a torus we
have \be C_s^{\rm op} = 1. \ee Can a similar optimization be
obtained for regular color codes? The answer is yes, and the
corresponding lattice is illustrated in
Fig.~\ref{figura_regulares}(d). For this code we have \be C_c^{\rm
op} = \frac{9}{8}, \ee very close to the value for surface codes.
Again, attaching $l^2$ copies of this lattice together gives a
family of codes of distance $4l$ with the same ratio $n/d^2$.
Therefore, we conclude that in a torus color codes encode twice as
qubits but surface codes require slightly less physical qubits for
optimal regular lattices.

Is this true also for other surfaces or is it something particular
of the torus? Instead of trying to answer this question in general,
we consider which are probably the most important example of
topological stabilizer codes in practice. For this we mean planar
codes, that is, topological codes that can be placed in a piece of
planar surface. More particularly, we want to compare surface and
color codes encoding a single-qubit, which are the most interesting
not only as quantum memories but also for quantum computation
\cite{dennis_etal02}\cite{tetraUQC}.

The trick to obtain planar codes from lattices related to surfaces
without borders is the same for surface and color codes. In
particular, it is enough to remove plaquettes from the original
lattice until the resulting surface can be unfolded onto a plane.
The new lattice has borders, and we have to explain which are the
strings that play now the role that closed strings played before in
the case of a compact surface like the torus. First, consider what
happens when a dark plaquette, that is, the corresponding plaquette
operator, is removed from a surface code. Take any $Z$-string that
has an endpoint in the removed plaquette. Since this $Z$-string
operator does not commute with the plaquette operator from the
removed plaquette, prior to the removal it was not in $\mathcal Z$,
but after the removal it could be, at least as long as this endpoint
is concerned. Therefore, the removal of a dark (light) plaquette
creates a dark (light) border in which only $Z$-strings
($X$-strings) can end at. Additionally, some string operators
winding around the removed plaquette are no longer boundaries, but
these considerations do not have relevance in the geometries that we
are considering. As for color codes, the situation is similar. When
a blue plaquette is removed, a blue border is created in which only
blue strings can end, and so on and so forth.

\begin{figure}
 \includegraphics[width=8cm]{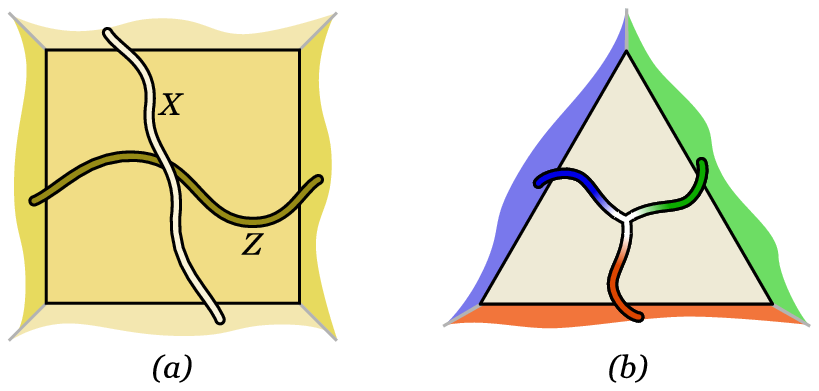}
 \caption{
 The geometry of planar codes encoding a single qubit. The colors in
 the borders represent the class of the missing face. Only suitable
 strings can have endpoints at each border. The string operators
 corresponding to encoded operators are shown. (a) A planar
 surface code. (b)
 The triangular (color) code, in which encoded operators
 are related to a string-net.
 }
 \label{figura_planar}
\end{figure}

With these ideas in mind, it is not difficult to understand the code
geometries shown in Fig.~\ref{figura_planar}, which solves the
question we have raised before.

For surface codes there are two borders of each type, so that the
nontrivial $X$-strings ($Z$-strings) connect light (dark) borders,
see Fig.~\ref{figura_planar}(a). In the case of color codes, there
are three borders, one of each color, and the nature on the encoded
operators shows a feature of color codes which we have not discussed
yet. The point is that in color codes string-nets are allowed. In
particular, branching points in which three different colored
strings of the same type meet are allowed. This means that such a
configuration does not violate any of the plaquette conditions. Then
for these triangular codes, the encoded $\bar{X}$ and $\bar{Z}$
operators are constructed with such a string-net, as
Fig.~\ref{figura_planar}(b) shows.
\begin{figure}
 \includegraphics[width=8cm]{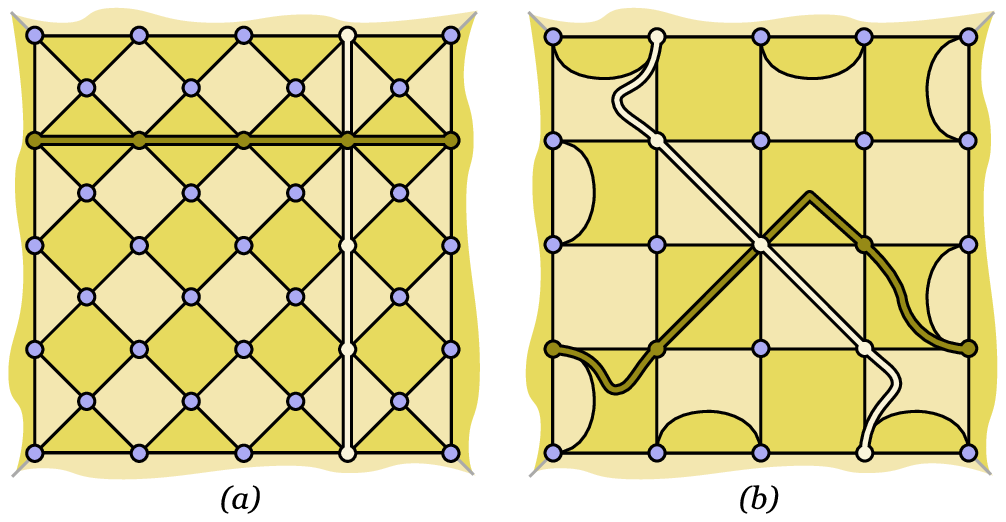}
 \caption{
 Surface codes encoding a single qubit and with distance $d=5$. (a)
 A non-optimal version with $n=41$. (b) An optimized version with
 $n=25$.
 }
 \label{figura_singlesurface}
\end{figure}

We want to consider the efficiency of these families of planar
codes, which encode a single-qubit. In
Fig.\ref{figura_singlesurface} two different versions of the surface
code with distance $d=5$ are shown. The (a) version belongs to the
original family of single-qubit surface codes, for which the
asymptotic value of the ratio is \be C_s \sim 2. \ee This is not the
best that can be done, as the code (b) shows with approximately half
the number of qubits and equal distance. In fact, the optimized
value for planar surface codes is \be C_s^{\rm op}=1. \ee

\begin{figure}
 \includegraphics[width=8cm]{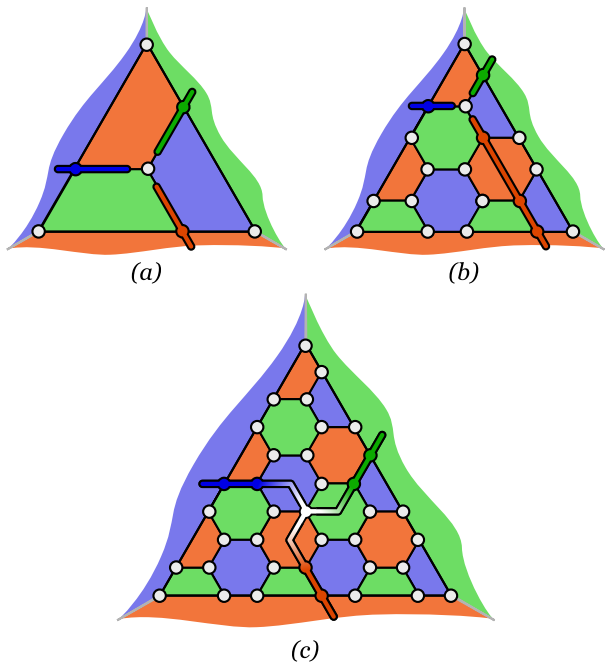}
 \caption{
 Color codes encoding a single qubit. (a) Triangular code with
 distance $d=3$ and number of qubits $n=7$. (b) Triangular code with
 $d=5$ and $n=19$. (c) Triangular code with $d=7$ and $n=37$.
 }
 \label{figura_triangular}
\end{figure}

As for triangular codes, examples for distances $d=3,5,7$ are shown
in Fig.\ref{figura_triangular}. It is straightforward to continue
this family for arbitrarily large distances. For these color codes
the asymptotic value yields the following optimized value \be
C_c^{\rm op}\sim\frac{3}{4}. \ee

Therefore, triangular
codes are not only particularly interesting for quantum
computation \cite{tetraUQC}, but also more efficient in terms of the number of
physical qubits required.

Finally, we would like to touch upon how topological
stabilizer codes give rise to the idea of self-protected quantum
memories. To this end, one must consider a physical system in
which qubits are placed according to the geometry of the
topological code, and introduce certain Hamiltonian dictated by
the generators of the stabilizer. In the case of surface codes,
the Hamiltonian is
\be
H:= -\sum_{p\in P_D} B_p^X - \sum_{p\in P_L} B_p^Z,
\ee
while for color codes it is
\be
H:= - \sum_{p} \left( B_p^X + B_p^Z\right).
\ee
One of the main differences between these Hamiltonians is that for
color codes all plaquettes play the same role, whereas in the
case of surface codes we have to distinguish between light and
dark plaquettes. In any case, in both cases the ground states
correspond to encoded states, and there exists a gap which
separates them from excited states. Moreover, no local operator
can connect ground states. Only those operators with a
topologically nontrivial support are able to distinguish among
these protected states, something which makes this quantum
memories remarkably robust against perturbations with a local
nature.

\section{Conclusions}
\label{sect_conclusions}

In this paper we have made a presentation of surface codes
\cite{kitaev97} and color codes \cite{topologicalclifford} on equal
footing. This allows us to make a comparative study of their
properties in more detail, such as the possible set of gates that
they can implement transversally and the number $k$ of encoded
qubits (logical qubits).

We have also introduced the notion of error correcting rate
$C:=\frac{n}{d^2}$ as a means to evaluate the performance of
topological 2D codes as far as error correction capabilities is
concerned. We have computed this figure of merit for surface codes
and color codes in the most representative and important topologies:
the torus and the plane. In the torus, we find that the optimal
value for surface codes is $C_s=1$, while for color codes we find
$C_c=\frac{9}{8}$, which is are very close. For practical
applications, planar codes are the most valuable topologies. For
them we find that the optimized values for surface codes are again
$C_s=1$, but this time color codes yield a better value
$C_c=\frac{3}{4}$. Having in mind that the number of encoded logical
qubits for color codes is always, i.e. in any topology, twice as
much as for surface codes \eqref{encoding}, this means that color
codes demand less resources.

 \noindent {\em Acknowledgements} We
acknowledge financial support from a PFI fellowship of the EJ-GV
(H.B.), DGS grant  under contract BFM 2003-05316-C02-01 (M.A.MD.),
and CAM-UCM grant under ref. 910758.



\begin{thebibliography}{99}
\bibitem{tetravalent}
A tetravalent lattice is a lattice with coordination number four,
i.e., each site has four nearest neighbours sites.
\end{thebibliography}
\end{document}